\documentstyle[aps,epsf,twocolumn]{revtex}

\def\be{\begin{equation}}
\def\ee{\end{equation}}
\def\bea{\begin{eqnarray}}
\def\eea{\end{eqnarray}}

\def\t12h{\frac{\theta_{12}}{2}}

\def\eps{\varepsilon}

\def\r#1{(\ref{#1})}
\def\nn{\nonumber\\}

\def\app{{a^{\prime\prime}}}
\def\cpp{{c^{\prime\prime}}}
\def\NPB#1#2#3{{ Nucl. Phys.} {B{\bf #1}} #2 (#3)}
\def\PRD#1#2#3{{ Phys. Rev.} {D{\bf #1}} #2 (#3)}
\def\PRB#1#2#3{{ Phys. Rev.} {B{\bf #1}} #2 (#3)}
\def\PRL#1#2#3{{ Phys. Rev. Lett.} {\bf #1} #2 (#3)}
\def\TMP#1#2#3{{ Theor. Math. Phys.} {\bf #1} #2 (#3)}
\def\CMP#1#2#3{{ Comm. Math. Phys.} {\bf #1} #2 (#3)}
\def\IJMPA#1#2#3{{ Int. Jour. Mod. Phys.} {A{\bf  #1}} #2 (#3)}

\begin{document}


\preprint{OUTP-98-21S}

\title{\bf Sine-Gordon low-energy effective theory for Copper Benzoate}
\author {Fabian H.L. E\char'31ler}
\address{Department of Physics, Theoretical Physics,
        Oxford University\\ 1 Keble Road, Oxford OX1 3NP, United Kingdom}

\maketitle
\begin{abstract}
Specific heat data for the quasi one
dimensional quantum magnet Copper Benzoate (${\rm
Cu(C_6D_5COO)_2\cdot3D_2O}$) is analyzed in the framework of an
effective low-energy description in terms of a Sine-Gordon theory. 
\end{abstract}

\begin{narrowtext}
\section{Introduction}
Quasi one dimensional quantum magnets have been a focus of intense
theoretical and experimental interest for a long time. Most of the
work is based on and centers around the spin-$1/2$ Heisenberg model
\cite{heisenberg}.

Starting with Bethe's seminal work \cite{bethe} a host of exact
results have been obtained for ground state properties \cite{hulthen},
magnetic susceptibility \cite{yaya}, thermodynamics \cite{tak,klumper2},
excitation spectrum \cite{ft,exmagn} and correlation functions
\cite{lp,corr1,corr2,corr0}. From the point of view of standard spin
wave theory, which is highly successful for ``three dimensional''
materials, the findings of the these investigations were rather
unusual. Over the last thirty years a number of anisotropic materials
have been found that constitute excellent realizations of the one dimensional
Heisenberg model \cite{materials}, {\sl e.g.} ${\rm KCuF_3}$, ${\rm
Sr_2CuO_3}$, ${\rm Cs_2CuCl_4}$ or CuPzN, and many theoretical predictions
have been confirmed experimentally. One main focus of attention
was the spectrum, which comprises of an incoherent (two
particle) scattering continuum of elementary excitations, the
so-called spinons \cite{ft}. These can be visualized in terms of
ferromagnetic ``domain walls'' and are strikingly different from the usual
spin waves. In particular spinons are believed to have fractional
(semionic) exclusion statistics \cite{frac}. The low-energy effective
theory of the spin-$1/2$ chain is simply a free massless boson
\cite{lp,vladb,luk}
\be
{\cal L}=\frac{1}{2}\left(\partial_\mu\Phi\right)^2\ .
\label{free}
\ee

It has been known for some time that Copper Benzoate is another
realization of a quasi-1D $S=1/2$ Heisenberg antiferromagnet
\cite{date}. However, its response to a magnetic field has been
found to be unusual \cite{oshima}: structural anisotropy leads to to
generation of small staggered fields in and perpendicular to the
direction of the applied field. Early specific heat measurements
\cite{takeda} showed behaviour incompatible with theoretical results
for a simple Heisenberg chain. 

In a series of recent experiments \cite{magn,dender} the behaviour of
Copper Benzoate in a magnetic field was investigated in great detail. 
Neutron scattering experiments \cite{dender} established the existence
of field-dependent incommensurate low energy modes. The
incommensurability was found to be consistent with the one predicted
by the exact solution of the Heisenberg model in a magnetic
field. However, the system exhibited an unexpected excitation gap
induced by the applied field. As no evidence for ordering was found
in the experiments down to temperatures of $0.1-0.3K$ the interchain
coupling in Copper Benzoate supposedly is very small (the exchange is
approximately 18K). We therefore will neglect it in the present work.

In \cite{oa} it was proposed that Copper Benzoate is described
by the Hamiltonian
\be
H_{\rm CuB}=\sum_i J \vec{S}_i\cdot\vec{S}_{i+1}- {\tt g}\mu_B H S_i^z + 
\mu_B h (-1)^i S^x_i\ ,
\label{hamil}
\ee
where $H\gg h$, ${\tt g}$ is the effective Land\'e g-factor and
$J=1.57\ {\rm meV}$ \cite{magn}. Here the induced staggered field $h$
is a function of the known (staggered) $g$-tensor \cite{oshima} and
the Dzyaloshinskii-Moriya (DM) interaction in Copper Benzoate, for
which unfortunately only scant information is available. If direction
and magnitude of the DM interaction are given, $h$ is calculated as
follows \cite{oa}. The $g$-tensor in the $a'', b, c''$ basis (these
denote the three principal axes of the exchange interaction
\cite{foot0}) is given by \cite{oshima} 
\begin{equation}
 g = \left( \begin{array}{ccc}
              2.115 & \pm 0.0190 &  0.0906 \\
             \pm 0.0190 & 2.059 & \pm 0.0495 \\
             0.0906 & \pm 0.0495 & 2.316
            \end{array} \right).
\label{gtensor}
\end{equation}
The $\pm$ correspond to the two inequivalent sites and indicate that
application of a uniform field induces a staggered one. The
corresponding contribution to the Hamitonian is
\be
{\cal H}_{\rm magn}=g_{ab}\ \mu_B\ H_a\ S^b\ .
\label{hmag}
\ee
The staggered DM interaction is
\bea
{\cal H}_{\rm DM} &=& \sum_j (-1)^j\
\vec{D}\cdot(\vec{S}_j\times\vec{S}_{j+1})\, 
\label{hdm}
\eea
where $|\vec{D}|\ll J$ and the direction of $\vec{D}$ is thought to be
close to the $a^{\prime\prime}$ axis. The DM interaction is eliminated
by a rotation in spin space around the $\vec{D}$-axis by an angle
$\alpha=\pm(\arctan D/J)/2$ on even/odd sites. This induces a very
small exchange anisotropy which is negligible, and a staggered field
\be
{\cal H}_{\rm OA}=
\frac{g}{2}\mu_B(\vec{H}\times\vec{D}/J)\cdot\sum_j(-1)^j\vec{S}_j\ , 
\label{hoa}
\ee
where we used that $|\vec{D}|/J\ll 1$. Combining the two contributions
\r{hoa} and \r{hmag} we obtain the total induced staggered field.
For example, a uniform field applied along the $a^{\prime\prime}$ axis
induces a staggered field $h$ along the b axis of magnitude
$(0.019-2.115 D_c/2J) H$. 

The low energy effective theory of \r{hamil} is obtained by abelian
bosonization and is given by a Sine-Gordon model with Lagrangian
density \cite{oa}
\be
{\cal L} = \frac{1}{2} (\partial_\mu\Phi)^2 + \lambda(h) \cos(\beta
\ \Theta)\ .
\label{lagr}
\ee
Here $\Theta$ is the dual field and the coupling $\beta$ depends on
the value of the applied uniform field. The coefficient $\lambda(h)$
can at present not be calculated exactly. The reason is that the
amplitudes of the bosonized expressions of lattice spin operators for
$H>0$ are not known (in the absence of a magnetic field they have been
determined very recently in \cite{luk,luk2,affleck}).
For later convenience we define the
coupling
\be
\xi=\frac{\beta^2}{8\pi-\beta^2}\ .
\ee
The Sine-Gordon theory \r{lagr}, for all its apparent simplicity, has
fascinated physicists for decades. It is of interest as an integrable
classical nonlinear differential equation featuring soliton
solutions. On the quantum level it has been one of the cornerstones of
nonperturbative quantum field theory with many exciting features such
as quantum solitons, topological charge or regularization dependence
in the nonperturbative regime \cite{SG1}. Most importantly the quantum
Sine-Gordon model is exactly solvable \cite{SG} and many physically
important quantities can be calculated. In particular, the spectrum
is known to consist of a soliton-antisoliton doublet of mass $M$ and
their bound states which are called ``breathers''. 

The soliton mass gap $M$ can in principle be calculated exactly
\cite{zam} in terms of $\beta$ and $\lambda$ for a given
short-distance normalization of correlation functions. 
However, $\lambda$ is known only for the case of vanishing uniform
field $H=0$ (see above). A simple analysis based on the results of
\cite{zam,luk,affleck} yields for this case (i.e. one takes to
staggered field into account, but bosonizes at $H=0$)
\bea
M/J &\approx& \frac{2\Gamma(1/6)}{\sqrt{\pi}\Gamma(2/3)}
\left[\pi\frac{\Gamma(3/4)}{\Gamma(1/4)}\sqrt{\frac{\pi}{4(2\pi)^{3/2}}}
\right]^{2/3}\left(\frac{h}{J}\right)^{2/3}\nn
&\approx& 1.8 \left(\frac{h}{J}\right)^{2/3} \ ,
\eea
where we have neglected logarithmic corrections. This is in good
agreement with the numerical analysis of the lattice Hamiltonian
\r{hamil} for $H=0$, which gives \cite{oa}
\be
M/J \approx 1.85 \left(\frac{h}{J}\right)^{2/3}
\bigg|\log\frac{h}{J}\bigg|^{1/6}\ .
\label{solitonmass}
\ee
In addition to soliton and antisoliton there are $[\frac{1}{\xi}]$
(here $[\ ]$ is the integer part) breathers with masses 
\be
M_n = 2M\ \sin n\pi\xi/2\ ;\  n=1,\ldots ,\left[1/\xi\right]. 
\label{mass}
\ee
The mass spectrum as a function of magnetic field for Copper Benzoate
was explicitly determined in \cite{et}.

Exact predictions of the low-energy effective theory \r{lagr} for the
spectrum \cite{oa} and the dynamical structure factor \cite{et} were
found to be consistent with Neutron scattering experiments for
applied fields along the b-axis. 
It is interesting to note that the Sine-Gordon solitons and breathers
are fundamentally different from the spinons of the spin-$1/2$ chain. 

In \cite{dender} precise measurements of the low-temperature specific
heat were presented and analyzed in terms of several noninteracting
one-dimensional bosons of the same mass. On the other hand,
the spectrum of the Sine-Gordon model in the relevant region of
couplings features five interacting modes with different masses
\cite{et} (soliton, antisoliton and three breathers).

In the present work we analyze the specific heat data of \cite{dender}
in the framework of the Sine-Gordon theory.
A very important input in the low-energy effective Lagrangian \r{lagr}
are the values of the coupling $\beta$ and the spin velocity $v_s$. In
a ``minimal'' model (MM) they are calculated from the exact Bethe
Ansatz solution of the Heisenberg XXX chain in an applied magnetic
field $H$ \cite{vladb}. In Appendix A we summarize the corresponding
relevant Bethe Ansatz results.
\begin{figure}[ht]
\begin{center}
\noindent
\epsfxsize=0.45\textwidth
\epsfbox{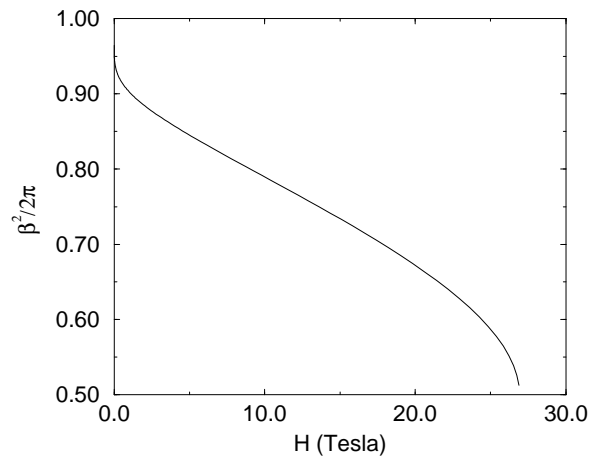}
\end{center}
\caption{\label{fig:beta}%
Coupling constant $\beta^2/2\pi$ in the MM as a function of the
applied magnetic field $H\parallel b$
.} 
\end{figure}

\begin{figure}[ht]
\begin{center}
\noindent
\epsfxsize=0.45\textwidth
\epsfbox{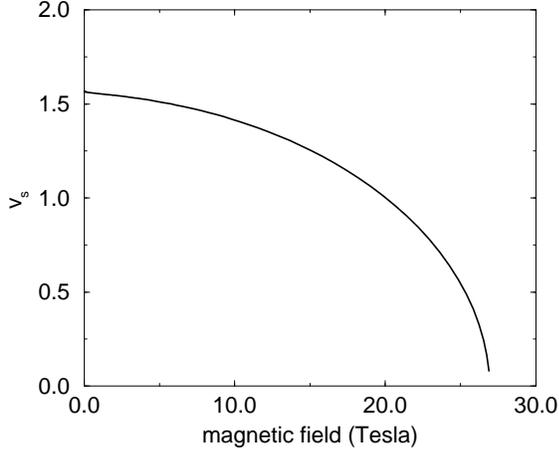}
\end{center}
\caption{\label{fig:vf}%
Spin velocity in the MM as a function of the applied field $H\parallel
b$ 
.} 
\end{figure}

This procedure appears to be reasonable as long as
the induced field $h$ is very small so that its effects on $\beta$ and
$v_s$ are negligible. The results are shown in Figs~\ref{fig:beta} and
\ref{fig:vf} respectively. At very high fields $g \mu_B H\approx 2J$
we approach the incommensurate-commensurate transition to the
saturated ferromagnetic state \cite{JNPT} and the spin velocity thus
tends to zero. 

An alternative scenario is to consider the spin velocity and/or the
coupling $\beta$ as a phenomenological parameters in the Sine-Gordon
theory \r{lagr}. The rationale behind such an approach is that the
known presence of the DM interaction as well as possible exchange
anisotropies will lead to deviations in the values of these quantities
as compared to the MM predictions. 

A simple calculation shows that the effect of the DM interaction on
$\beta$ and $v_s$ is negligible. Adding an exchange anisotropy 
\be
H_{\rm anis} = -J\Delta\sum_j S^x_j S^x_{j+1}\ , 0<\Delta\ll 1
\label{anis}
\ee
to the Hamiltonian \r{hamil} firstly induces a change in the spin
velocity entering the effective Lagrangian \r{lagr} and secondly
generates the second harmonic of the SG interaction, {\sl i.e.} the
effective low-energy theory becomes \cite{foot3}
\be
{\cal L} = \frac{1}{2} (\partial_\mu\Phi)^2 + \lambda \cos(\beta
\ \Theta)\ + \mu \cos(2\beta \ \Theta)\ .
\label{lagr2}
\ee
Here we have assumed that $\Delta$ is much smaller than the magnetic
energy scale $g\mu_BH$.
The coupling  $\mu$ mainly depends on $\Delta$ (the second harmonic
also gets generated at 1-loop level by the $\cos\beta\Theta$
interaction). In the regime of couplings $\beta$ we
are interested in, the second harmonic is a relevant operator (in the
RG sense), although it is of course much less relevant than
$\cos\beta\Phi$. This means that we can safely neglect the second
harmonic, {\sl unless $\lambda\ll\mu$ }! We will return to this point
below. 
Physically the effect of \r{anis} is the following: if no uniform
field is applied, the system remains critical. The spin velocity
and the critical exponents are changed slightly. If a uniform field is
applied perpendicular to the direction of the anisotropy, a spectral
gap forms, even if no staggered field is generated. 

\section{Sine-Gordon Thermodynamics}

The thermodynamics of the Sine-Gordon model is most efficiently
studied \cite{ddv} {\sl via} the recently developed Thermal Bethe
Ansatz approach \cite{tba}, which circumvents problems associated with
solving the infinite number of coupled nonlinear integral equations
that emerge in the standard approach based on the string hypothesis
\cite{stringthermo} (note that the coupling constant $\beta$ in our
problem is a continously varying quantity and no truncation to a
finite number of coupled equations is possible). It was shown in
\cite{ddv} that the free energy of the Sine-Gordon model can be
expressed in terms of the solution of a single nonlinear integral
equation for the complex quantity $\eps(\theta)$ (we set the spin
velocity to 1 for simplicity)
\bea 
&&\eps(\theta) = -i M \beta\sinh(\theta+i\eta^\prime) \nn
&&- \int_{-\infty}^\infty d\theta^\prime
G_0(\theta-\theta^\prime) 
\ln\left(1+\exp\left[-\eps(\theta^\prime)\right]\right)\nn
&&+\int_{-\infty}^\infty d\theta^\prime
G_0(\theta-\theta^\prime+2i\eta^\prime) 
\ln\left(1+\exp\left[-\bar{\eps}(\theta^\prime)\right]\right),
\label{inteq}
\eea
where $\beta=(k_BT)^{-1}$, $M$ is the soliton mass and
\bea
G_0(\theta) &=&\int_0^\infty \frac{d\omega }{\pi^2}
\frac{\cos(2\omega\theta/\pi)
\sinh(\omega(\xi-1))}{\sinh(\omega\xi)\cosh(\omega)}\ .
\eea

The free energy density is given by

\bea
f(\beta)&=&-\frac{M}{\beta\pi}\!\Im m\!\int_{-\infty}^\infty\!\!\!\!\!\!
 d\theta\ 
\sinh(\theta+i\eta^\prime) \ln\left[1+e^{-\eps(\theta)}\right].
\eea

As we are interested in the attractive regime $\gamma<4\pi$ we have
\be
0<\eta^\prime<\pi\xi/2\ .
\label{etap}
\ee
Note that the free energy does not depend on the value of
$\eta^\prime$ as long as it is chosen in the interval \r{etap}.
The set \r{inteq} of two coupled nonlinear integral equations is
solved by iteration. For $\beta\to\infty$ the first iterations can be
calculated analytically and the corresponding contributions to the
free energy are seen to be of the form
\bea
f(\beta)&\sim& -\frac{2M}{\beta\pi}\sum_{n=1}^\infty 
\frac{(-1)^{n+1}}{n}K_1(nm\beta)\nn
&&-\frac{M_1}{\beta\pi}K_1(M_1\beta)+\ldots\ ,
\eea
where $M_1=2M\sin\frac{\pi\xi}{2}$ is the mass of the first breather
and $K_1$ is a modified Bessel function. The first term is the
contribution of soliton-antisoliton scattering states to the free
energy, whereas the second term is the contribution of the first
breather. Both terms have the form characteristic of massive
relativistic bosons. The contributions of the heavier breathers are
found in higher orders of the iterative procedure employed in solving
\r{inteq}. The specific heat is obtained from the free energy
\be
C=T\frac{\partial^2f(\beta)}{\partial T^2}\ .
\ee
At low temperatures it is found to be of the form
\bea
C&\sim& \sum_{\alpha=1}^{[1/\xi]}
\frac{k_B}{\sqrt{2\pi}v_s}\left[1+\frac{k_BT}{M_\alpha}+
\frac{3}{4}\left(\frac{k_BT}{M_\alpha}\right)^2\right]\nn
&&\times
\left(\frac{M_\alpha}{k_BT}\right)^{3/2}\exp(-M_\alpha/k_BT)\ ,
\eea
where $M_\alpha$ are given by \r{mass}. In order to compare
theoretical predictions based on the SGM with the specific heat data
of \cite{dender} we need the free energy at ``intermediate''
temperatures and thus have to resort to a numerical solution of
\r{inteq} by iteration. 

\section{Specific Heat in Copper Benzoate}

Let us now investigate the question how well the theoretical
predictions based on an effective Sine-Gordon theory agree with the
specific heat data of \cite{dender}. As was pointed out in
\cite{dender}, at very low temperatures a nuclear contribution to the
specific heat is present. In the following analysis we neglect this
contribution but note, that by taking it into account we can achieve
excellent agreement of the SG results with the data at low
temperatues. 

We now analyze the specific heat data of \cite{dender}
as follows: we calculate the specific heat in the framework of the
MM using the soliton gap as a free parameter, which is then fixed by
fitting the calculated specific heat to the data. This procedure
yields the dependence of the gap on the applied field $M(H)$, which
has to be consistent with \r{solitonmass} and the dependence of $h$
and $H$, which follows from \r{hmag} and \r{hoa}.
In order to keep things simple we ignore the logarithmic correction
and $H$-dependence of $\beta$ in \r{solitonmass} so that
\be
M(H)\approx c H^{2/3}\ .
\label{scaling}
\ee
Here the coefficient $c$ depends on the orientation of the applied
field and the direction and magnitude of the DM interaction as
explained above. In order to calculate $c$, we would need to know the
precise magnitude as well as orientation of the DM interaction as is
clear from \r{hoa}. Unfortunately this information is presently not
available. From considerations based on the crystal structure
$\vec{D}$ is expected to lie in the
$a^{\prime\prime}-c^{\prime\prime}$ plane and and is thought to be
roughly of magnitude $D/J\sim .1$.

Reversing the logic \cite{oa}, if we determine the coefficient $c$ in
\r{scaling} for all three independent orientations of the applied
field by fitting the SGM predictions to the data, we can calculate
what the direction and magnitude of $\vec{D}$ has to be in order to
reproduce these results. Below we follow this line of argument.

\subsection{Magnetic Field along $c^{\prime\prime}$ axis}

For magnetic fields along the $c^{\prime\prime}$ axis we find
excellent agreement of the data with the ``minimal'' model discussed
above. This is shown for some values of $H$ in Figs \ref{fig:caxis1}
and \ref{fig:caxis2}.

\begin{figure}[ht]
\noindent
\epsfxsize=0.45\textwidth
\epsfbox{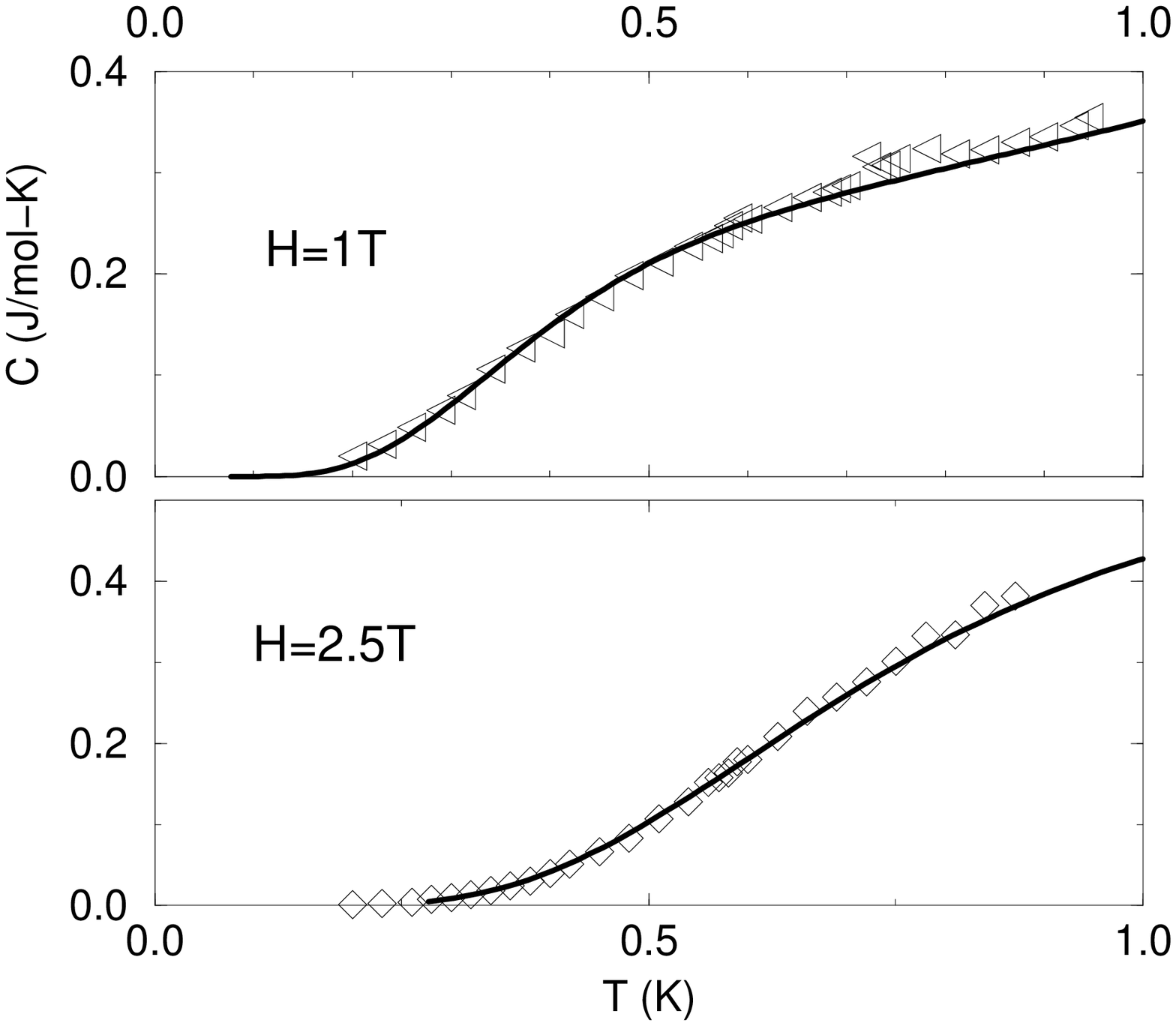}
\caption{\label{fig:caxis1}%
Specific heat as a function of temperature for fields of
H=1T and H=2.5T applied along the $c^{\prime\prime}$ axis
.}
\end{figure}

As explained above, the presence of an exchange anisotropy would
change the spin velocity. Assuming $v_s$ to be eight percent smaller
than in the MM we still obtain good agreement with the data (note that
the soliton mass is of course changed as well) as is shown in
Fig. \ref{fig:caxis3}. 

\begin{figure}[ht]
\noindent
\epsfxsize=0.45\textwidth
\epsfbox{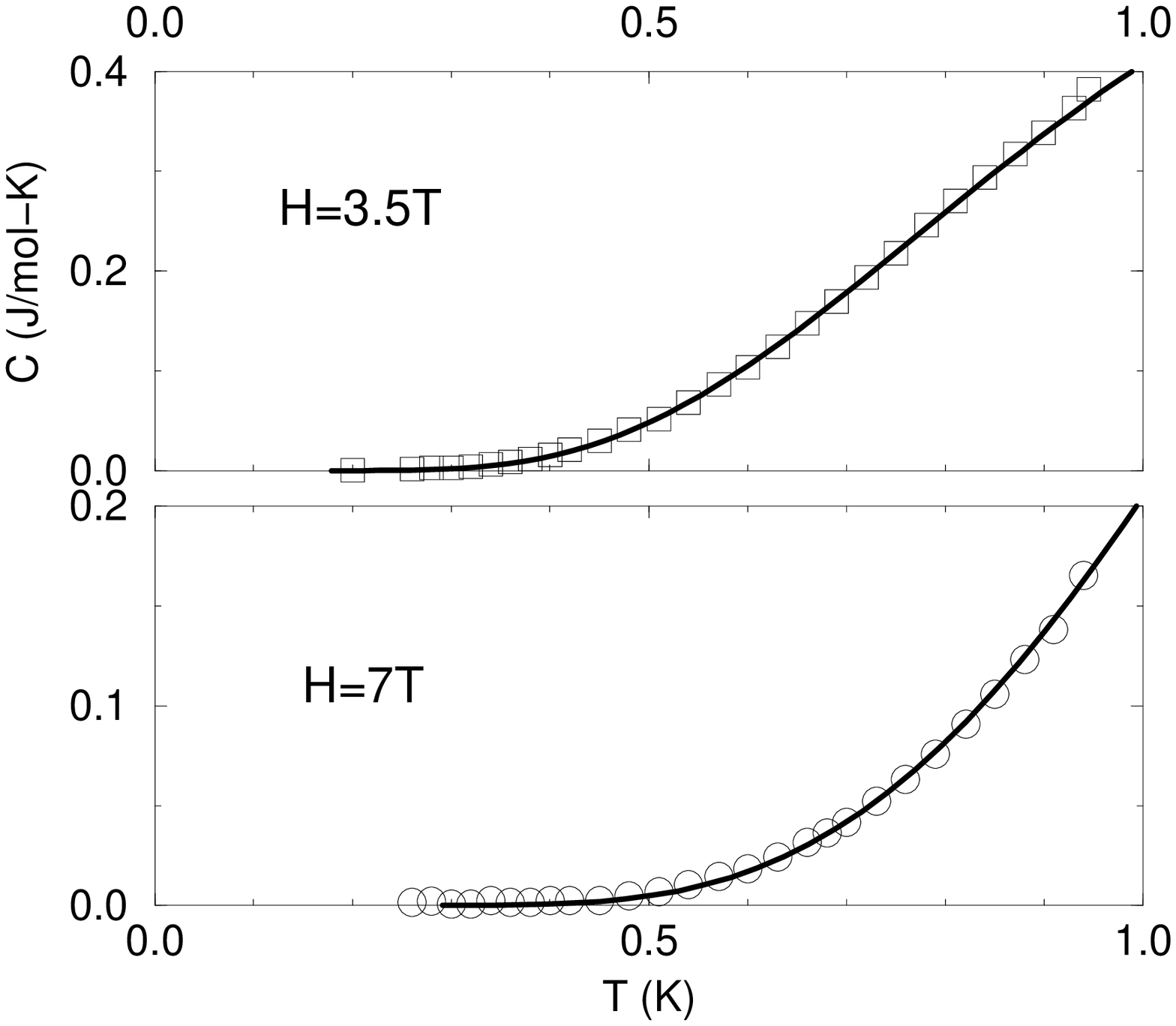}
\caption{\label{fig:caxis2}%
Specific heat as a function of temperature for fields of H=3.5T and
H=7T applied along the $c^{\prime\prime}$ axis 
.}
\end{figure}

\begin{figure}[ht]
\noindent
\epsfxsize=0.45\textwidth
\epsfbox{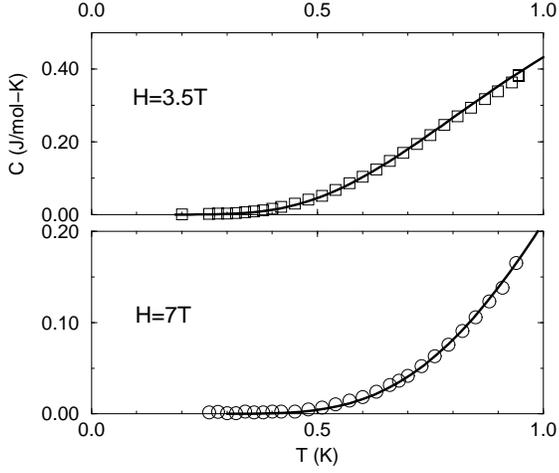}
\caption{\label{fig:caxis3}%
Specific heat as a function of temperature for fields of
H=3.5T and H=7T applied along the $c^{\prime\prime}$ axis. The spin 
velocity is taken to be eight percent smaller than in the MM
.}
\end{figure}
In order to check which scenario for $v_s$ is in better agreement with
experiment the values for the soliton masses obtained by fitting to
the data have to be compared with \r{solitonmass}.

\subsection{Magnetic Field along $b$ axis}
For magnetic fields along the b-axis the agreement of the MM
prediction with the data is less impressive. As is clear from 
Figs~\ref{fig:baxis1}-\ref{fig:baxis2} the MM systematically underestimates
the measured specific heat in the temperature region $T\approx 0.8-1K$
although there is still fair agreement of the MM with experiment.

\begin{figure}[ht]
\noindent
\epsfxsize=0.45\textwidth
\epsfbox{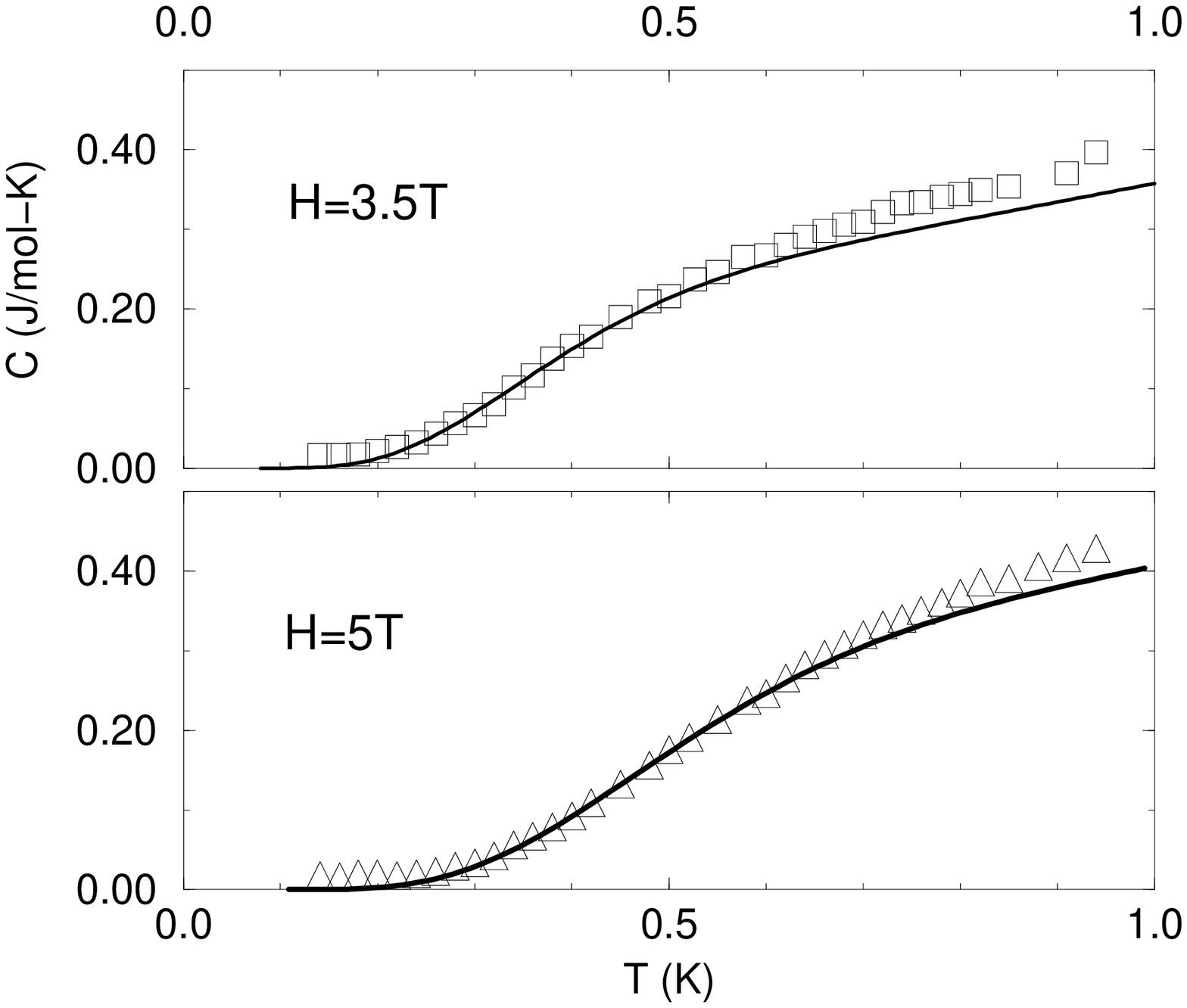}
\caption{\label{fig:baxis1}%
Specific heat as a function of temperature for fields applied along
the $b$ axis
.}
\end{figure}

\begin{figure}[ht]
\noindent
\epsfxsize=0.45\textwidth
\epsfbox{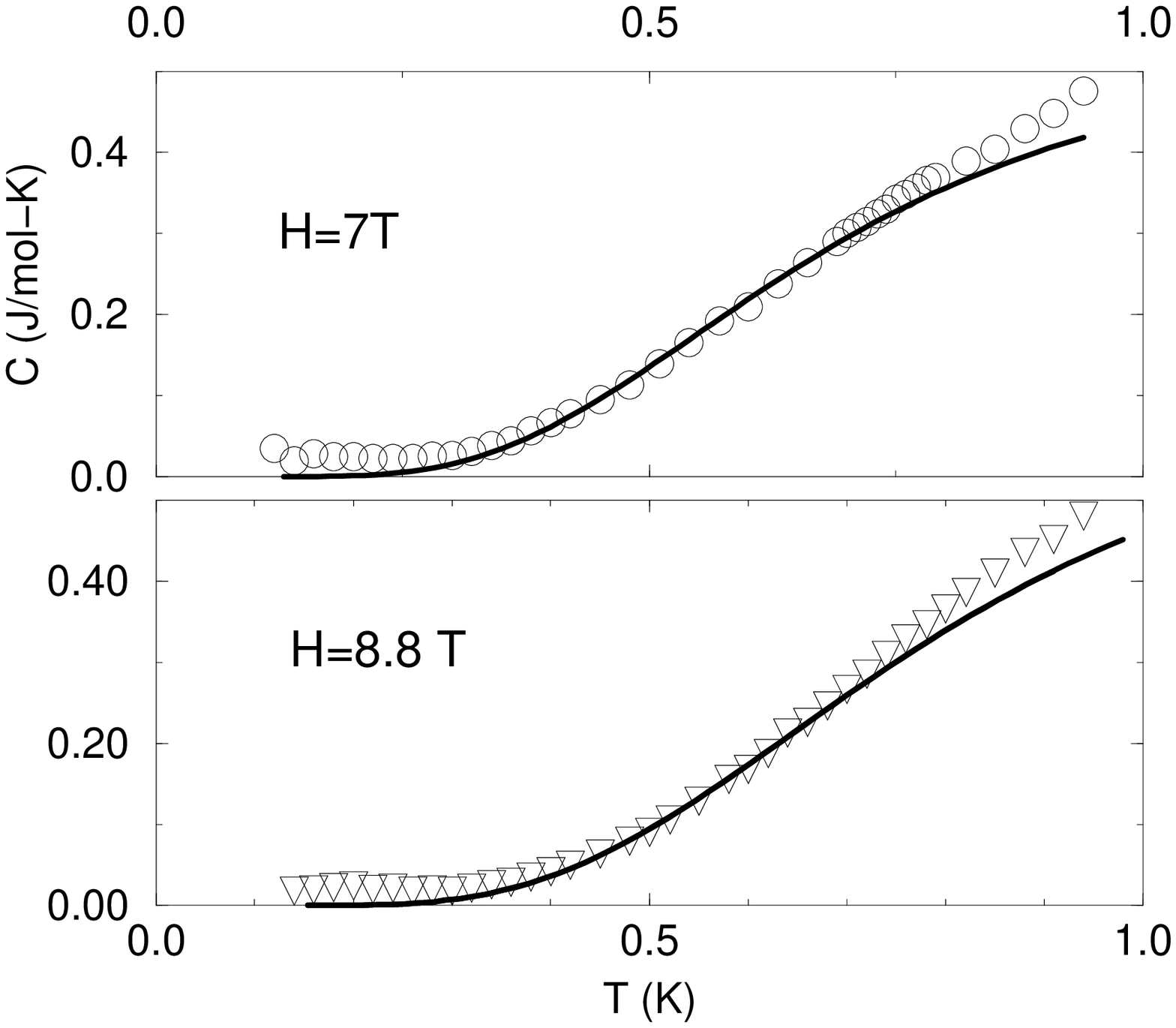}
\caption{\label{fig:baxis2}%
Specific heat as a function of temperature for fields applied along
the $b$ axis
.}
\end{figure}

A much improved fit to the data is obtained if the spin velocities in
the effective SGM are changed by eight percent as compared to the
MM. This is shown in Figs~\ref{fig:baxis3}-\ref{fig:baxis4}. 

\begin{figure}[ht]
\noindent
\epsfxsize=0.45\textwidth
\epsfbox{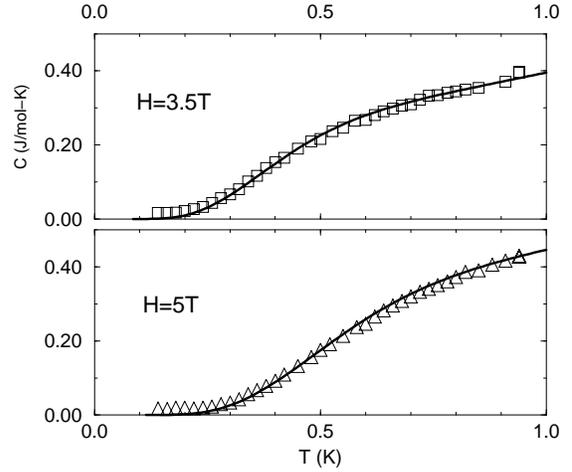}
\caption{\label{fig:baxis3}%
Specific heat for fields applied along the $b$ axis. 
The spin velocity is taken to be eight percent smaller than in the MM
.}
\end{figure}

\begin{figure}[ht]
\noindent
\epsfxsize=0.45\textwidth
\epsfbox{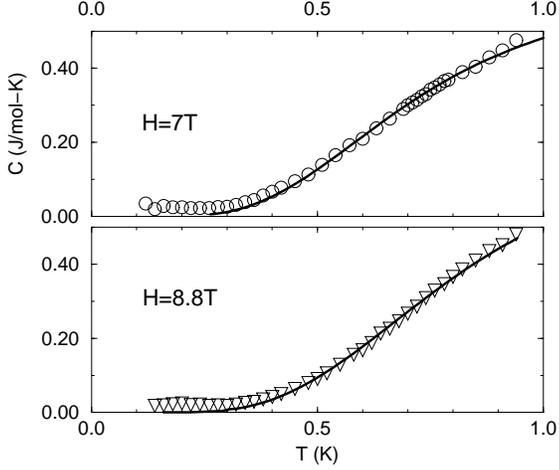}
\caption{\label{fig:baxis4}%
Specific heat for fields applied along the $b$ axis. 
The spin velocity is taken to be eight percent smaller than in the MM
.}
\end{figure}

In order to check the compatibility of the fitted values for the
soliton gap $M_{\rm fit}(H)$ with \r{scaling} we plot $M_{\rm fit}(H)$
as a function of $H$ in Fig.~\ref{fig:gap}. For simplicity we only
consider the results calculated on the basis of the MM. We find good
agreement for applied fields along the $b$ and $c^{\prime\prime}$
axes. The logarithmic corrections \r{solitonmass} to the gap may
improve the agreement, but need (unavailable) information
on the DM interaction as input. The ratio of mass gaps for fields
applied along the $b$ and $\cpp$ axes is found to be
\be
\Delta_\cpp/\Delta_b\approx 1.43/0.65=2.2\ .
\ee

\begin{figure}[ht]
\noindent
\epsfxsize=0.45\textwidth
\epsfbox{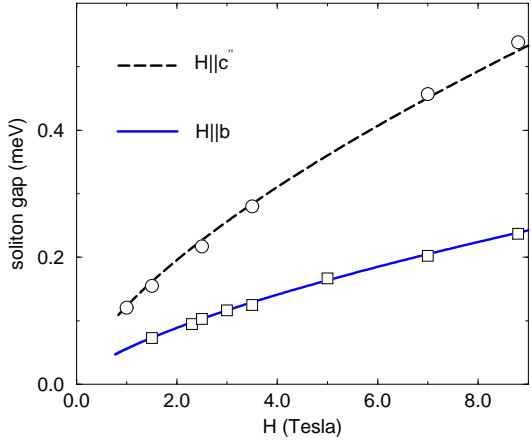}
\caption{\label{fig:gap}%
Gap of the soliton for $H\parallel c^{\prime\prime}$ and
$H\parallel b$. The fits are to the simple scaling law $M\propto
(H/J)^{2/3}$ as described in the text
.}
\end{figure}

\subsection{Magnetic Field along $a^{\prime\prime}$ axis}

For fields applied along the $a^{\prime\prime}$ axis it is impossible
to obtain agreement of the MM predictions with the data. How can we
understand this fact? The DM interaction is expected to lie in the
$a^{\prime\prime}-c^{\prime\prime}$ plane, so that we can write
\be
\vec{D}=D_a \vec{e}_{a^{\prime\prime}}+D_c \vec{e}_{c^{\prime\prime}}\
.
\ee
For $\vec{H}=H\vec{e}_{a^{\prime\prime}}$ the net induced staggered
field is directed along the $b$ axis and is of magnitude
\be
h=(0.019-2.115 D_c/2J)H\ .
\ee
Clearly this would be very small if $D_c/J\approx 0.02$. We note that
such a value of $D_c$ together with the gap ratio
$\Delta_{\cpp}/\Delta_b\approx 2.2$ implies that $D_a/J\approx 0.12$.
Note that this is consistent with the expectation that $D/J\approx .1$
and a direction of $\vec{D}$ close to the $\app$ axis. 

In this case the coupling $\lambda(h)$ in \r{lagr} would become very
small and there would be a regime in which perturbations other than
the staggered field would dominate the physics. For example, if an
exchange anisotropy \r{anis} was present in Copper Benzoate, the
low-energy effective theory would be given by
\r{lagr2} with $\mu\gg\lambda$. As a first approximation we then can
ignore the $\lambda\cos\beta\Theta$ term and study the remaining
(repulsive) SGM. In the framework of this scenario we obtain
a rather reasonable fit to the data as is shown in
Fig.~\ref{fig:aaxis}. The expected mass gap is
difficult to estimate, because as we already mentioned the
coefficients in the bosonization formulas are known only in the
absence of a magnectic field \cite{luk,affleck,luk2}. A crude estimate
can be obtained by approximating the coefficents in the presence of a
uniform field by the ones for $H=0$. The gaps obtained by fitting the
data are found to be consistent with a rather small anisotropy
$\Delta/J{<\atop \approx} 0.05$ in this approximation. We note in
passing that the zero-field specific heat found in \cite{dender}
\be
C(T)=0.68(1)R k_BT/J\ ,
\ee
actually corresponds to an anisotropy of the type \r{anis} with
$\Delta/J\approx 0.06$.

In order to work out a more quantitative theory
for the $\app$-axis specific heat data the full two-frequency
Sine-Gordon theory would need to be analyzed, which is possible in a
perturbative framework \cite{ffpt}. We hope to address this point in
the future. The (non) existence of the mechanism described above
could in principle be checked by inelastic Neutron scattering
with $\vec{H}\parallel\vec{e}_\app$: if the physics is indeed
dominated by interactions other than the induced staggered field, the
spectrum will be very different from the one oberved in
\cite{dender}. In particular, if exchange anisotropy is the relevant
mechanism and the effective low-energy theory is thus given by
\r{lagr2} with $\lambda\approx 0$, then no coherent one-particle
excitations are present. The dynamical structure factor at wave number
$\pi$ is then dominated by an incoherent soliton-antisoliton
continuum.

\begin{figure}[ht]
\noindent
\epsfxsize=0.45\textwidth
\epsfbox{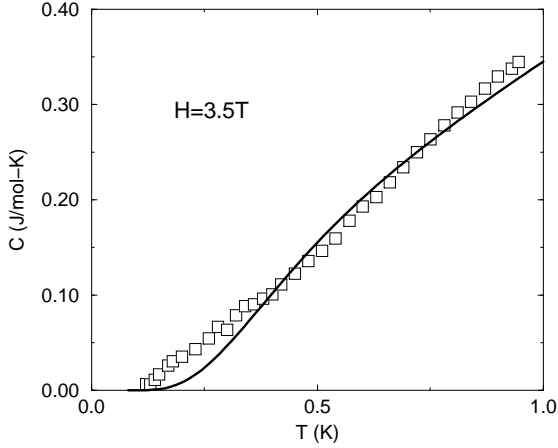}
\caption{\label{fig:aaxis}%
Specific heat for $H\parallel a^{\prime\prime}$. The theoretical curve
is obtained by assuming an exchange anisotropy as explained in the text
.}
\end{figure}

\section{Conclusions}
We have analyzed specific heat data for
Copper Benzoate in the framework of a Sine-Gordon low-energy effective
theory. For uniform magnetic fields applied along the $b$ and $\cpp$
axes we find good agreement of the theory with the specific heat
data. The $\app$ axis data cannot be understood by the same theory
that applies for the $b$ and $\cpp$ axes. We argue that the staggered
field induced by the DM interaction essentially cancels the field
induced by the staggered g tensor for $\vec{H}\parallel
\vec{e}_{\app}$ so that a new mechanism is responsible for generating
the gap. We propose that exchange anisotropy might be responsible.

\begin{center}
{\bf Acknowledgements}
\end{center}
I am very grateful to D. Reich for generously providing the
experimental data and to D. Dender for many helpful explanations
concerning Copper Benzoate. I also would like to thank J. Chalker and
A.M. Tsvelik for important discussions and the EPSRC for support via
an Advanced Fellowship.

\vskip .5cm
\appendix
\begin{center}
{\bf APPENDIX A: THE HEISENBERG CHAIN IN A FIELD}
\end{center}
\vskip .5cm
We summarize some relevant results (for derivations see \cite{vladb})
for the anisotropic Heisenberg model in a magnetic field
\be
H_{\rm XXZ}= -4J\sum_n \vec{S}_n\cdot\vec{S}_{n+1}+(\Delta -1)
S^z_nS^z_{n+1} -2hS^z\ ,
\label{hxxz}
\ee
where $-1\leq\Delta=\cos\gamma<1$ and $J>0$. The customary form of the
Hamiltonian is obtained by performing the unitary transformation 
\be
S_n^z\longrightarrow S_n^z\ ,\qquad S_n^\pm\longrightarrow (-1)^n S_n^\pm\ .
\label{uni}
\ee
At low energies \r{hxxz} is described by a free massless boson
\r{free} compactified on a ring of radius $R$ i.e. $\Phi$ and
$\Phi+2\pi R$ are identified. The dual field $\Theta$ fulfils
$\Theta=\Theta+1/R$ (see e.g. \cite{eggert}). The following
bosonization rules can be derived along the lines of e.g. \cite{GNT}:
$\vec{S}_n \longrightarrow [\vec{ J}(x) + \vec{ n} (x)], \
x=na_0$, where $a_0$ is the lattice spacing and

\bea
J^z&=&\frac{a_0}{\beta}\partial_x\Phi(x)\ ,\nn 
(-1)^n J^+&=&i {\cal A}(\delta x) \exp\left(i\beta\Theta(x)\right)\
a_0\partial_x\Theta(x)\nn 
&&-\frac{{\cal B}}{2\pi} \exp\left( -i\beta\Theta(x)\right)\
\sin\left(\frac{2\pi}{\beta}\Phi(x)-2\delta x\right),\nn 
n^x (x) &=& {\Lambda_\perp}{} \cos(\beta\Theta (x)),\
n^y (x) = {\Lambda_\perp}{} \sin(\beta\Theta (x)),\nn
n^z (x) &=& -(-1)^n{\Lambda_z}{} \sin(\frac{2\pi}{\beta}\Phi
(x)-2\delta x).
\label{boso}
\eea
Here $\beta=2\pi R$ and the coefficients $\Lambda_{\perp,z}$ are
known only in the absence of a magnetic field \cite{luk,luk2}. 
The standard structure in terms of uniform and staggered magnetization
operators is obtained by performing the unitary transformation \r{uni}.
We note that the often neglected first term in $J^+$ is actually more
important than the second: as a matter of fact, for 
$H=0$ it yields the leading contribution to transverse correlations at
wave number $\pi$ of \r{hxxz} \cite{luk,foot1}.
For $\beta=\sqrt{2\pi}$ i.e. the SU(2) invariant chain in zero field,
the second term is simply the sum of left and right SU(2)
currents. The first contribution to $J^+$ corresponds to a
particle-hole excitation with spin 1 relative to the ground state (see
(II.5.3) and (XVIII.1.16) of \cite{vladb}) and for $H=0$ can be
derived by carefully taking the continuum limit of the Jordan-Wigner
lattice fermions \cite{affleck2,luk}. For $H\neq 0$ we expect ${\cal
A}(\delta x)$ on general grounds to be of the form ${\rm const}\times
\cos(2\delta x)$. Equations \r{boso} are used to derive
the continuum form of the perturbation \r{anis} (note that both smooth
and staggered components of the spin operators contribute for
$\beta=\sqrt{2\pi}$). 

The constant $\beta$ and spin velocity $v_s$ are determined by $\Delta$,
$J$ and $h$ of the lattice model as follows. The dressed energy
$\eps$, momentum $p$ and ``charge'' $Z$ of an elementary spinon are
given in terms of the solutions of the linear integral equations

\bea
\eps(\lambda)&-&\int_{-A}^A \frac{d\mu}{2\pi}\
K(\lambda-\mu)\ \eps(\mu) = 2h-\frac{4J\sin^2\gamma}{\cosh 2\lambda
+\cos\gamma},\nn
p(\lambda)&=&2\pi\int_0^\lambda d\mu\ \rho(\mu)\ ,\nn
\rho(\lambda)&-&\int_{-A}^A \frac{d\mu}{2\pi}\
K(\lambda-\mu)\ \rho(\mu) = \frac{2\sin\gamma}{2\pi[\cosh 2\lambda
+\cos\gamma]}\ ,\nn 
Z(\lambda)&-&\int_{-A}^A \frac{d\mu}{2\pi}\
K(\lambda-\mu)\ Z(\mu) = 1\ ,
\label{inteqs}
\eea
where $K(\lambda)=2\sin 2\gamma/(\cosh\lambda -\cos 2\gamma)$.
Here $A$ is the rapidity corresponding to the Fermi momentum and is
fixed by the condition
\be
\eps(\pm A)=0\ .
\ee
The spin velocity is then given by the derivative of the spinon energy
with respect to the momentum at the Fermi surface
\be
v_s=\frac{\partial\epsilon(\lambda)}{\partial
p(\lambda)}\bigg|_{\lambda=A}=\frac{\partial\epsilon(\lambda)/\partial\lambda}
{2\pi\rho(\lambda)}\bigg|_{\lambda=A} \ .
\ee
Finally, $\beta$ and $\delta$ are given by
\be
\beta= \frac{\sqrt{\pi}}{Z}\ , \
\delta=\frac{\pi}{2}-\pi\int_{-A}^A d\mu\ \rho(\mu)\ .
\ee
In order to determine $v_s$ and $\beta$ we solve \r{inteqs}
numerically, which is easily done to very high precision as the
equations are linear. The results are shown in Figs~\ref{fig:beta} and
\ref{fig:vf}. 
Finally, we note that correlation functions at small finite
temperatures can be calculated like in \cite{corr1} (see also
\cite{gia}). We only must remember to shift the momentum by $\pm
2\delta$ away from $\pi$ for the longitudinal correlation function and
use the correlation exponent as calculated above from the Bethe
Ansatz. For example the transverse dynamical susceptibility at small
momentum (which corresponds to momentum $\pi$ in the customary form of
the Heisenberg Hamiltonian, which is related to \r{hxxz} by the
unitary transformation \r{uni}) is given by
\bea
\chi_\perp(\omega, q)&\propto& T^{-2+\beta^2/2\pi}\ 
B(\frac{\beta^2}{8\pi}-i \frac{\omega-v_s q]}{4\pi T},
1-\frac{\beta^2}{4\pi})\nn
&&\times B(\frac{\beta^2}{8\pi}-i \frac{\omega+v_s q}{4\pi T},
1-\frac{\beta^2}{4\pi})\ ,
\eea
where $B(x,y)$ is the Beta function and $q$ is close to zero.
The longitudinal susceptibility is dominated by the gapless modes at
$\pi\pm 2\delta$. It is the sum of two terms 
\bea
\chi_\parallel(\omega, q)&\propto& \sum_{\sigma=\pm}
T^{-2+2\pi/\beta^2}\ 
B(\frac{\pi}{2\beta^2}-i \frac{\omega-v_s Q_\sigma}{4\pi T},
1-\frac{\pi}{\beta^2})\nn
&&\times B(\frac{\pi}{2\beta^2}-i \frac{\omega+v_s Q_\sigma}{4\pi T},
1-\frac{\pi}{\beta^2})\ ,
\eea
where $Q_\sigma= q-\pi+\sigma 2\delta$. 
\end{narrowtext}


\begin{thebibliography}{99}
\bibitem{foot0}
The definition of the principal axes goes back to the ESR analysis of
\cite{date}. The corresponding exchange anisotropies are expected to
be small (in fact too small to give rise to the gap induced by the
applied field \cite{oa}), so that the isotropic Hamiltonian \r{hamil}
is a natural starting point for theoretical studies of the
field-induced gap.
\bibitem{foot3}
We set the velocity equal to one in order to simplify all
equations. The velocity dependence is easily restored by dimensional
analysis.
\end{thebibliography}
\end{document}